\documentclass{jpsj-suppl}
\usepackage{txfonts} 
\providecommand{\e}{$e^{\scriptscriptstyle-}$}
\providecommand{\ep}{$e^{\scriptscriptstyle-}\!/\!p^{\scriptscriptstyle+}$}
\newcommand{\RB}{\textbf{R}$\times$\textbf{B}{} }

\title{High Precision Experiments with Cold and Ultra-Cold Neutrons }

\author{Hartmut \textsc{Abele}$^{1}$, Tobias \textsc{Jenke}$^{1}$, Erwin \textsc{Jericha}$^{1}$, Gertrud \textsc{Konrad}$^{1}$, Bastian \textsc{M{\"a}rkisch}$^{2}$, Christian \textsc{Plonka}$^{3}$, Ulrich \textsc{Schmidt}$^{2}$, Torsten \textsc{Soldner}$^{4}$. }

\inst{$^{1}$Atominstitut, Technische Universit{\"a}t Wien, Stadionallee 2, 1020 Wien, Austria \\
$^{2}$Physikalisches Institut, Universit{\"a}t Heidelberg, Im Neuenheimer Feld 226, 69120 Heidelberg, Germany \\
$^{3}$Physikalisches Institut, Universit{\"a}t Mainz, Staudingerweg 7, 55128 Mainz, Germany \\
$^{4}$Institut Laue-Langevin, 71 Avenue des Martyrs, 38000 Grenoble, France}

\email{abele@ati.ac.at}

\recdate{October 14, 2014}

\abst{This work presents selected results from the first round of the DFG Priority Programme SPP 1491 "precision experiments in particle and astroparticle physics with cold and ultra-cold neutrons".
}

\kword{Standard model, gravitation, charge quantization, neutron decay, parity violation,  CKM-matrix}

\begin{document}
\maketitle

\section{Introduction}

New high intensity sources for ultra-cold neutrons are coming into operation having the potential to exceed contemporary source strengths by several orders of magnitude.  This priority programme wants to exploit these new technologies and implement novel concepts as a source of neutron decay products. It addresses some of the unsolved questions of modern science: the nature of the fundamental forces and underlying symmetries, as well as the nature of the gravitational force at very small distances.  New facilities and technological developments now open the window for significant improvement in precision by 1-2 orders of magnitude.  This allows to probe these questions in a complementary way to LHC-based experiments or even constitutes a unique way.  The research program focuses on four \textbf{priority areas}, which are directly related to specific physics/astrophysics issues:

\begin{itemize}
	\item \textbf{Priority Area A}
CP-symmetry violation and particle physics in the early universe (addressed mainly by the search for the neutron electric dipole moment)

	\item \textbf{Priority Area B}
The structure and nature of weak interaction and possible extensions of the Standard Model (addressed mainly by precise studies of the neutron $\beta$-decay)

	\item \textbf{Priority Area C}
Relation between gravitation and quantum theory (probed by investigations of low-energy bound states in the gravitational field)

	\item \textbf{Priority Area D}
Charge quantization and the electric neutrality of the neutron (probed by a precision test of the neutron's electric charge)
\end{itemize}
The intended improvement in experimental precision has to go in parallel with the development of new or improved measurement techniques which are often at the extreme border of feasibility.
\begin{itemize}
	\item \textbf{Priority Area E}
New techniques:
1) particle detection, 2) magnetometry, 3) neutron optics
\end{itemize}
This article concentrates on selected results of priority areas B, C, and D.  With these priority areas we aim for a cartography of the Standard Model of particle physics of the first particle generation including gravitation.

\section{Priority Area B}

Priority area B addresses a number of questions in particle physics, with main emphasis on the search for new physics beyond the Standard Model of particles physics, and in particular, on the question of unification of all forces shortly after the Big Bang.  This grand unification is not part of the Standard Model, and new symmetry concepts are needed like left-right symmetry, fundamental fermion compositeness, new particles, leptoquarks, supersymmetry, and many more.
In the search for new symmetries, we see experiments with cold and ultra-cold neutrons as part of the high-precision frontier in the domain of low-energy studies.  These experiments fit in a greater field of precision measurements comprising cold or ultra-cold neutrons, cold or ultra-cold ions or atoms, protons, electrons, and their antiparticles.  Precise symmetry tests of various kinds will be boosted by the proposed facility PERC~\cite{Kon11}.  Interesting are symmetry contributions to neutron $\beta$-decay parameters, which are forbidden in the Standard Model.  Projects using the PERC facility will test the Standard Model at a much higher level of sensitivity benefiting both, from the gain in statistical accuracy for individual measurements and from the redundancy of observables accessible.  Neutron decay offers a number of independent observables, considerably larger than the small number of parameters describing this decay in the Standard Model.  The first element $V_{\rm ud}$ of the quark mixing CKM matrix is one of the two parameters describing neutron decay within the Standard Model, the other being the ratio $\lambda$ = $g_{\rm A}/g_{\rm V}$ of axial-vector to vector coupling constants including their relative phase.
Observables in free neutron decay are abundant:  besides the lifetime $\tau_{\rm n}$, angular correlations involving the neutron spin as well as momenta and spins of the emitted particles are characterized by individual coefficients, which can be related to the underlying coupling strengths of the weak interaction.  Examples are the electron-antineutrino correlation coefficient $a$~\cite{Str77,Byr02,Bae08}, the beta asymmetry parameter $A$~\cite{Abe97,Abe02,Mun13,Pla12,Men13}, the neutrino asymmetry parameter $B$~\cite{Kre05a,Sch07} (reconstructed from proton and electron momenta), the proton asymmetry parameter $C$~\cite{Sch08}, the triple correlation coefficient $D$~\cite{Sol04,Chu12}, the Fierz interference term $b$, and various correlation coefficients involving the electron spin~\cite{Koz09,Koz12}.  Each coefficient in turn relates to an underlying broken symmetry.  A method of loss-free spectroscopy is presented in Ref.~\cite{Abe93}.  So far, the measurement of Schumann et al.~\cite{Sch08} is the only precision measurement of $C$ (see also Ref.~\cite{PDG14}), with a 1.1\,\% error: 0.4\,\% statistics and 1\,\% systematics. Many of the correlations are yet unobserved.

\noindent  Within priority area B, we want to include the following precision tests of the Standard Model and search for physics beyond the Standard Model with the new instrument PERC:
\begin{itemize}
\setlength{\itemsep}{-3pt}

	\item  An improved determination of Standard Model parameters.  With a new and precise value of $\lambda$, we will cover the demand from particle and astroparticle physics.  The derived value for $V_{\rm ud}$ gives better insight into quark mixing.	
	
	\item  A search for right-handed admixtures to the left-handed feature of the Standard Model.  As a natural consequence of symmetry breaking in the early universe, they should be found in neutron $\beta$-decay.  Signatures are a $W_{\rm R}$ mass with mixing angle $\zeta$.	
	
	\item  A search for scalar and tensor admixtures $g_{\rm S}$ and $g_{\rm T}$ to the electroweak interaction.  $g_{\rm S}$ and $g_{\rm T}$ are also forbidden in the Standard Model but supersymmetry contributions to correlation coefficients can approach the 10$^{-3}$ level, a factor of five away from the current sensitivity limit~\cite{Sch07}.	
	
	\item  A first search in neutron $\beta$-decay for the Fierz interference term $b$, which is forbidden in the Standard Model but can approach the 10$^{-3}$ level from supersymmetry contributions.	
	
	\item  A first measurement of the weak-magnetism form factor $f_2$ prediction of electroweak theory.  Such an experiment would be one of the rare occasions, where a strong test of the underlying structure itself of the Standard Model becomes available.
\end{itemize}

\subsection{Source of neutron decay products PERC}

\begin{figure}[tbh]
	\centering
	\includegraphics[scale=.3]{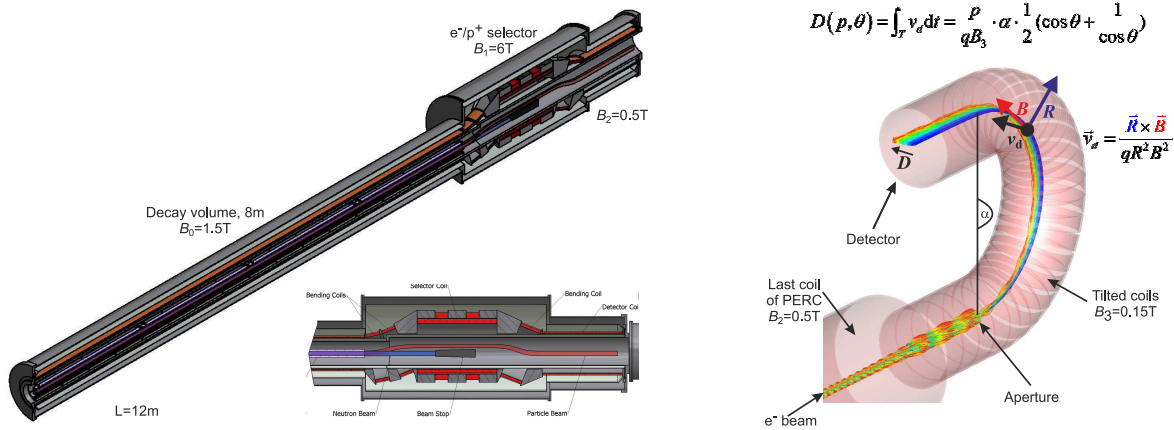}
	\caption{Left:  Preliminary design of the \ep-beam facility PERC~\cite{Kon12}, with a zoom on the \ep-selector in the lower part.  PERC consists, inter alia, of a 12\,m long superconducting magnet system that produces a strong magnetic field of 1.5-6\,T.  Right: Scheme of the \RB drift detector~\cite{Wan13} connected to the end of PERC, with simulated \e-trajectories.  The various colours correspond to the \e-momenta, from 0 (blue) to 1.19\,MeV/c (red).}
	\label{f1_2}
\end{figure}

The main goal within the first round of the priority programme was the design of the source of neutron decay products PERC (Proton and Electron Radiation Channel)~\cite{Dub08,Kon12}, shown in Fig.~\ref{f1_2} (Left), together with design and construction of the beam line at the Heinz Maier-Leibnitz Zentrum in Garching.  This work is a joint project of the Universities of Heidelberg and Mainz, Institut Laue-Langevin (ILL), Technische Universit{\"a}t M{\"u}nchen, and the Technische Universit{\"a}t Wien.  The first milestones were magnetic field calculations for strong magnetic fields.  They are needed for all three, particle transport, suppression of backscattering, and the determination of the effective solid angle for particle collection.  Now we are ready for the construction phase of this project.

\noindent At the exit, PERC delivers a quasi beam of decay electrons and protons (\ep) under well-defined and precisely variable conditions, which can well be separated from the cold neutron beam itself.  The next step is systematic studies for the analysis of the extracted electrons and protons.  Depending on the decay parameters studied this analysis must be performed with different and specialised detectors and secondary spectrometers.

\subsubsection{\RB drift spectrometer}

We have designed a new type of momentum spectrometer, shown in Fig.~\ref{f1_2} (Right). This spectrometer provides momentum analyses of decay electrons and protons at the PERC facility by using the \RB drift effect to disperse the charged particles~\cite{Wan13}. For that purpose a uniformly curved magnetic field is used. The spectrometer measures the particles with large phase space acceptance and high resolution.  Instead of eliminating the guiding field by a non-adiabatic transition, it is gradually adapted to the curved analysing field.  In the uniformly curved field, the charged particles can be adiabatically transported during dispersion and detection.  Indeed, the particles drifts have similar properties as their dispersion in normal magnetic spectrometers.  But, the \RB spectrometer is especially ideal for measurements of particles with low momenta and large incident angles.  For instance, particles with very small momentum can even be measured.  Besides, for particles with solid angle smaller than 10$^\circ$, the maximum aberration is less than 10$^{-4}$.  We reach a momentum resolution of 14.4\,keV/c, if the position resolution of the detector in the detection plane is better than 1\,mm.

\subsection{Polarisation on the 10$^{-4}$ level}

Next-generation precision measurements of asymmetries in neutron decay aim for $10^{-4}$ accuracy and require an adequate knowledge of the neutron beam polarisation.  In order to improve neutron polarisation and polarimetry by one order of magnitude, an opaque test bench (OTB)~\cite{Sol11,Kla13a} was installed at the instrument PF1B~\cite{Abe06} of the ILL.  The OTB consists of two opaque $^3$He spin filter cells with adiabatic fast passage (AFP) spin-flippers for the $^3$He spin.  The neutron wavelength is determined either by time-of-flight behind a chopper or by a neutron velocity selector.  Using the OTB, it was demonstrated experimentally that neutron polarisation can be measured with an accuracy of better than $10^{-4}$~\cite{Sol11}.  The efficiency of an AFP flipper for the neutron spin was measured to be higher than 99.99\,\%~\cite{Sol11}.  The polarisation of neutron beams with super mirror polarisers in X-SM geometry is limited by depolarisation in the polarising super mirrors~\cite{Kre05}.  In order to investigate and reduce this depolarisation, the OTB was used in a reflectometer geometry:  A beam with $>$99.99\,\% polarisation was reflected on a neutron mirror and analysed with the second opaque spin filter.  The mirror could be magnetised by a field of up to 0.8\,T.  These experiments demonstrated the dependence of depolarisation on material, magnetising field, super mirror factor (`$m$'), and angle of incidence~\cite{Kla12,Kla13,Kla13a}.

\subsection{Advanced flexible neutron beam tailoring}

In connection with the formation and preparation of the neutron beam into the PERC decay volume we are developing a novel spatial neutron spin resonator based on the Drabkin-type configuration with individually adressable resonator elements~\cite{Goe12,Goe13}.  This resonator allows us to tailor the spectral range and time structure of a polarised neutron beam in an extremely flexible and purely electronic way.  First successful experimental results have been obtained with thermal neutrons at the TRIGA reactor of the Atominstitut in Vienna and with very cold neutrons at the high-flux reactor of the ILL.  In contrast to the methods described in the preceeding paragraph, the
neutron wavelength is defined by a static magnetic selector field which
imposes a resonance condition on the incident neutron spectrum.

\section{Priority Area C}
 A theory of gravity has not been formulated within quantum field theory. String theory is one of the most promising theoretical concepts on this way. However, the price to pay are extra spatial dimensions which will have impact on the behaviour of the gravitational force at small, submillimeter distances. But also other ideas lead to modifications of Newton's gravitational law probed at very small distances.

The basic problem in searching for new physics at much smaller distances is that the size of the objects under study must be reduced, too, going along with a reduction of signal intensity. At the same time, the electrostatic background increases. A possible way out is the interaction of a macroscopic system (mirror for neutron reflection) with a pure quantum mechanical system (a neutron in the gravity potential of the earth). The proof of principle has been achieved with the observation of bound quantum states in the gravitational field in 2002~\cite{Nes02,Nes05,Wes07}, which triggered new experiments and activity in this direction~\cite{Abe08}. Here, the quantum states have pico-eV energy, compared with the typical energy scale for bound electron in an atom (eV), thus opening the way to a new technique for gravity experiments.

In priority area C, we explore the advantage of this unique system consisting of a particle, the neutron, and a macroscopic object, the mirror, i.e., the existence of quantum interference in quantum phases as observable and its precise measurement. In the newly proposed experiments, phase measurements in gravity potentials can be related to frequency measurements with unprecedented accuracy.
\subsection{The \emph{q}Bounce experiment}
\begin{figure}[t]
	\centering
	\parbox{0.45\textwidth}{
		\includegraphics[width=0.45\textwidth]{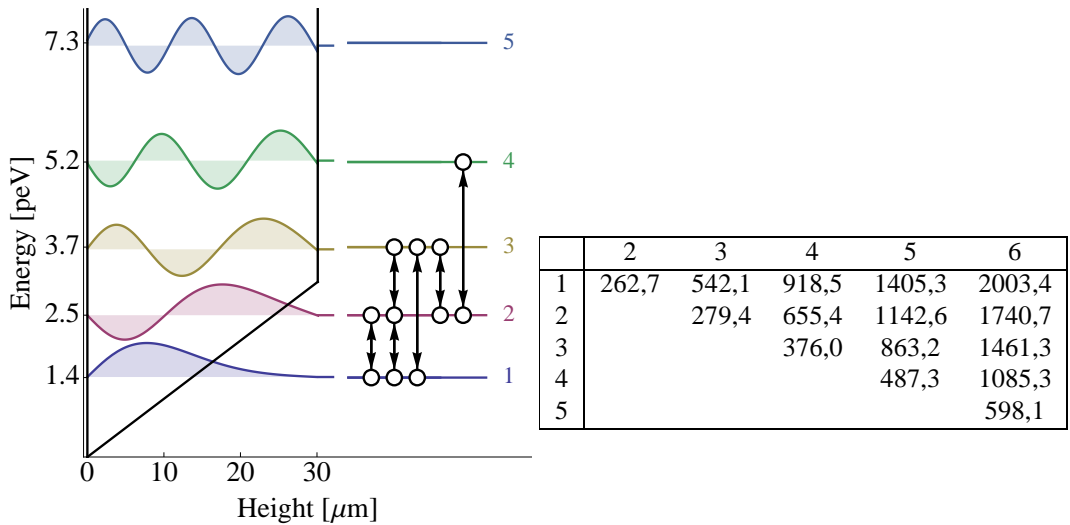}
	}
	\parbox{0.45\textwidth}{
		\begin{minipage}{0.45\textwidth}
			\vspace{1.6cm}
			\begin{tabular}{|l|ccccc|}
			\hline
			&2&3&4&5&6\\\hline
			1	&	262,7 & 542,1 & 918,5 & 1405,3  & 2003,4  \\
			2	&          & 279,4 & 655,4 & 1142,6  & 1740,7  \\
			3	&       &          & 376,0 & 863,2   & 1461,3  \\
			4	&       &       &          & 487,3   & 1085,3  \\
			5	&       &       &       &             & 598,1   \\
			\hline
			\end{tabular}
		\end{minipage}
	}
	\caption{Left: Frequency reference for gravity experiments based on energy eigenvalues of a neutron in the gravity potential of the Earth with confining mirrors at the bottom and at the top separated by 30.1~$\mu$ m. Also shown is the neutron density distributions for energy levels $|1\rangle$ to $|5\rangle$. The Gravity Resonance Spectroscopy technique allows to drive transitions between these states. Quantum transitions $|1\rangle \leftrightarrow|2\rangle$, $|1\rangle \leftrightarrow|3\rangle$,  $|2\rangle \leftrightarrow|3\rangle$, and $|2\rangle \leftrightarrow|4\rangle$ have been observed.\newline
Right: Resonant transition frequencies in Hz at width $l = 30.1$~$\mu$ m between mirrors, see Methods.}
	\label{fig:En}
\end{figure}
The \emph{q}Bounce experiment focuses on the control and understanding of a gravitationally interacting elementary quantum system using the techniques of resonance spectroscopy. It offers a new way of looking at gravitation based on quantum technology: an ultra-cold neutron, a quantum particle, as an object and as a tool.

The ultra-cold neutron as an object gives access to all parameters describing Newtonian gravitation: the mass and the distance. What is more, the Einstein equation of General Relativity has on the right hand side the energy-momentum tensor. Vacuum energy or Einstein's cosmological constant are part of this tensor. In a modern interpretation, a scalar field, which might be responsible for the accelerated expansion of the universe, is mimicking such a vacuum energy, also called dark energy. With a hypothetical neutron to scalar field coupling, there are good prospects for a discovery of such a scalar field with neutrons. Ultra-cold neutrons provide already best limits on a scalar field coupling and they have the potential to find a signal for a scalars or exclude them completely~\cite{Jen14}. Another interesting example for a hypothetical force is provided by the axion, a pseudoscalar boson, in the previously experimentally unaccessible so-called astrophysical axion window. Axions are - together with neutrinos - dark matter candidates for the un-known mass fraction in the Universe. Best direct limits for an axion providing a spin mass coupling at short distances are provided by the \emph{q}Bounce experiment~\cite{Jen14}.

The ultra-cold neutron as a tool to resonance spectroscopy guaranties highest precision. It is insensitive many systematic effects limiting experiments in  the past, such as electromagnetic ones plagued by other quantum objects like atoms or ions. The use of neutrons as test
particles bypasses the electromagnetic background induced
by van der Waals and Casimir forces and other polarizability
effects. It has a long lifetime of about 880 s and high sensitivity compared to other elementary systems like positronium of muonium.

The key technique is a newly-developed Gravity Resonance Spectroscopy (GRS) method~\cite{Jen11}. It is named in that way, because the energy difference between quantum states in the gravity potential has a one-to-one correspondence to the frequency of a modulator, in analogy to the Nuclear Magnetic Resonance technique, where the zeeman energy splitting of a magnetic moment in an outer magnetic field is related to the frequency of a radio-frequency field. We expect a similar statistical sensitivity for small energy changes as measurements of the electric dipole moment~\cite{Bak06} since the same neutrons are used. The precision of our method relies on the fact that frequency measurements can be performed with incredibly high precision. Fig.~\ref{fig:En} shows the observed transitions between eigenstates and the Schr{\"o}dinger wave functions.

An other observable is the spatial density distribution of a free falling neutron above a reflecting mirror. The development of the position dependent neutron detectors makes it possible to visualize the square of the Schr{\"o}dinger wave function~\cite{Jen09,Abe09}. We have now at hand a high-precision gravitational neutron spectrometer with available spatial resolution of 1.5  $\mu$m. Neutrons are detected in CR39 track detectors after neutron capture in a coated $^{10}$Boron layer of 100 nm thickness. An etching technique makes the tracks with a length of about 3 $\mu$m to 6 $\mu$m visible~\cite{Jen13}.

This feature opens up a new avenue in quantum tests of the universality of the free fall. With the \emph{q}BOUNCE data we are in a position to derive both the inertial masse $m_i$ and the gravitational mass $m_g$ from the free fall of a neutron alone. The spatial modulation of a coherent superposition of the Schr{\"o}dinger wave functions is determined by the third root of the product of the two masses. The discrete energy spectrum of the resonant transitions depends on the inertial as well as the gravitational mass with different fractional powers. Knowing the local acceleration of the earth $g$ and measuring two observables, energy and spatial modulation, allows us for the first time, independently to determine the inertial mass of the neutron and its gravitational mass, and thus the gravitational mass of a single particle.


\section{Priority Area D}
The neutron charge has been searched for by neutron beam deflection in a strong electric field perpendicular to the direction of flight. A non-zero charge would yield a deflection of the neutron's path, detectable using a grating or a slit. The sensitivity depends on the resolution power of the experiment and therefore on parameters like the strength of the electric field and the grating dimensions and, thirdly, on the number of detected neutrons. The best limit on the neutron charge was obtained in an experiment twenty-five years ago using a strongly collimated cold neutron beam of high intensity and an optical imaging system consisting of curved mirrors and multi-slits, resulting into $q_n$ $<$ 1.8$\times$10$^{-21}$ $q_e$, (90 \% C.L.)~\cite{Bau88}.
Another experiment, with ultracold neutrons, was conducted nearly at the same time~\cite{Bor}. The lower intensity of the UCN beam was counterbalanced by the longer time, the slow UCN remained in the electric field region. The intrinsic discovery potential of this experiment was
$q_n$ = 3.6$\times$10$^{-20}$$q_e$ per day at the former UCN source of the Leningrad VVR-M reactor. However, the sensitivity was reduced by a factor of 2.5 due to misalignments of the setup and by the quality of the optical elements. During only three days of running this experiment produced the result
$q_n$ = -(4.3$\pm$7.1)$\times$ 10$^{-20}$ $q_e$. In the following sections we present two new approaches for neutron charge measurements.
\subsection{An optical device}

\begin{figure*}[t]
\centering{
\includegraphics[width=.5\textwidth, angle=0]{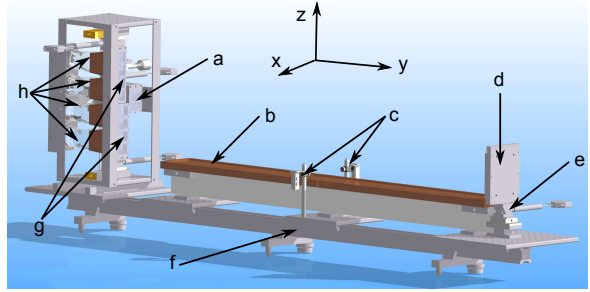}
\caption{\label{fig:Experiment} Draft of the experiment. a - input grating, b - liquid fomblin horizontal mirror, c - laser levels for alignment,
d - curved mirror, e - motorised goniometer, f - linear breadboard support, g - exit grating, h - detector.}
}
\end{figure*}

The first approach experiment uses an optical device and is performed at the UCN beamline of the UCN-facility PF2
at ILL. UCN are guided via a $\approx$1\,m long
stainless steel tube of 80\,mm diameter from the exit of the UCN turbine to the experiment.

Figure \ref{fig:Experiment} shows a scheme of the experiment; the $x,y,z$ directions are indicated;
$|z|$ is the direction of gravity. The vacuum is about $10^{-4}\,\rm mbar$ during all measurements.

After a rectangular entrance guide, the UCN pass an input grating (a). The grating has an overall dimension
of $45\times 50\,\rm mm^2$. The slits are 0.2\,mm wide with a pitch of 0.5\,mm. The lattice constant is therefore
0.6\,mm.
This entrance grating fragments the beam profile, and transmitted UCN are
reflected on a horizontal neutron mirror (b). The device can be described as an optical camera for ultra-cold neutrons:
A certain fraction of UCN reaches the end of the guide plates and encounters a mirror of cylindrical
shape (d) in the horizontal $x$-direction. The mirror has a
curvature of -1500\,mm with dimensions $100\times 150\,\rm mm^2$ in $x, z$-direction. It is mounted on a motorised goniometer (e) for alignment with laser levels (c).
It reflects the UCN and focuses the image of the entrance grating back onto
two stacked exit gratings (g). The gratings are made by laser cutting of a nickel foil. They have dimensions of 45\,mm in width and 145\,mm in height each. A detector array (h) is mounted behind the gratings. It detects the UCN passing the gratings.
Each grating can be shifted along the $x$-direction by a linear stage in order to scan the reflected beam image. Thus, one obtains a modulation in the count rate. At the points of the steepest slope, the sensitivity for deflections of the UCN beam is the highest.
By applying an electric field along the fligth path of the UCN, the charge of the free neutron can be determined.
On a beam time in 2013, a sensitivity of $\delta q_n  = 2.14 \cdot 10^{-20} \frac{e}{\sqrt{\mathrm{day}}}$ of this apparatus was obtained.
Further modifications like tuning the geometry, enhancing the electric field and reducing the slit width of the gratings should increase the sensitivity.
A first run has been taken place in November 2014.
\begin{figure}[tbh]
\centering
\includegraphics[scale=.3]{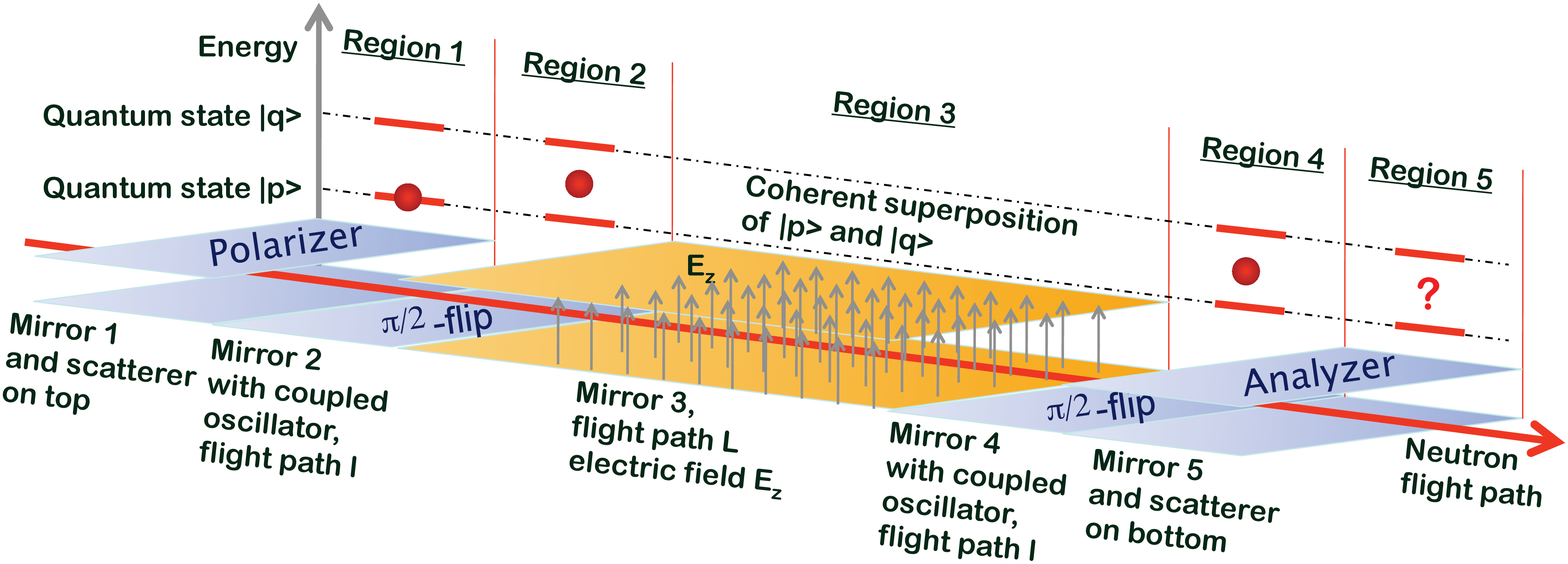}
\caption{Proposed experimental setup. Region 1: Preparation in a specific quantum state, e.g. state 1 with a polarizer.
Region 2: Application of first $\pi/2$ flip. Region 3: Flight path with length L. Region 4: Application of second $\pi/2$ flip. Region 5: State
analyzer; see also~\cite{Dur11}.}
\label{f6}
\end{figure}
\subsection{Probing the neutron's electric neutrality with Ramsey spectroscopy of gravitational
quantum states of ultracold neutrons}
This experiment uses an application of GRS to search for a charge of the neutron. The principle is explained in~\cite{Dur11}.  The method makes use of a Ramsey-technique -- see Fig.~\ref{f6}--  although until now it was considered to be impossible to use this technique for charge measurements, because a phase accumulation cancels by inverting the electrical field. But in the presence of an electric field $E_z$, the energy of quantum states in the gravity potential changes due to an additional electrostatic potential if a neutron carries a non-vanishing charge $q_n$. Important for this method is the fact that the energy shift differs from state to state due to the properties of a Schr{\"o}dinger wave packet in the linear potential of the earth.
It is of great advantage that the break down voltage and therefore the electric field available scales with the reciprocal square root of the distance.
With the already realized $E_z$ = 50 kV/mm, the discovery potential reads
 $\delta q_n  = 3 \cdot 10^{-20} \frac{e}{\sqrt{\mathrm{day}}}$.
Further improvements should increase the sensitivity.

This work in different research areas was supported by the Priority Programme SPP~1491 of the German DFG and the Austrian FWF.

\end{document}